# EFFECT OF PACKET DELAY VARIATION ON VIDEO/VOICE OVER DIFFSERV-MPLS IN IPV4/IPV6 NETWORKS

Md. Tariq Aziz[1], Mohammad Saiful Islam[2], Md. Nazmul Islam khan[3], and Prof. Adrian Popescu[4]

[1]Dept. of School of Computing, Blekinge Institute of Technology, Karlskrona, Sweden.
`tariq_ruet@yahoo.com`
[2]Dept. of School of Computing, Blekinge Institute of Technology, Karlskrona, Sweden.
`mdsaiful@gmail.com`
[3]Department of Electrical & Computer Engineering, Presidency University, Bangladesh
`nazmulk@mail.presidency.edu.bd`
[4]Dept. of School of Computing, Blekinge Institute of Technology, Karlskrona, Sweden.
`adrian.popescu@bth.se`

## ABSTRACT

*Over the last years, we have witnessed a rapid deployment of real-time applications on the Internet as well as many research works about Quality of Service (QoS), in particular IPv4 (Internet Protocol version 4). The inevitable exhaustion of the remaining IPv4 address pool has become progressively evident. As the evolution of Internet Protocol (IP) continues, the deployment of IPv6 QoS is underway. Today, there is limited experience in the deployment of QoS for IPv6 traffic in MPLS backbone networks in conjunction with DiffServ (Differentiated Services) support. DiffServ itself does not have the ability to control the traffic which has been taken for end-to-end path while a number of links of the path are congested. In contrast, MPLS Traffic Engineering (TE) is accomplished to control the traffic and can set up end-to-end routing path before data has been forwarded. From the evolution of IPv4 QoS solutions, we know that the integration of DiffServ and MPLS TE satisfies the guaranteed QoS requirement for real-time applications. This paper presents a QoS performance study of real-time applications such as voice and video conferencing in terms of Packet Delay Variation (PDV) over DiffServ with or without MPLS TE in IPv4/IPv6 networks using Optimized Network Engineering Tool (OPNET). We also study the interaction of Expedited Forwarding (EF), Assured Forwarding (AF) traffic aggregation, link congestion, as well as the effect of performance metric such as PDV. The effectiveness of DiffServ and MPLS TE integration in IPv4/IPv6 network is illustrated and analyzed. This paper shows that IPv6 experiences more PDV than their IPv4 counterparts.*

## KEYWORDS

*DiffServ, MPLS TE, IPv6, PDV and OPNET.*

## 1. INTRODUCTION

This Internet Protocol version 4 (IPv4) is one of the key foundations of the Internet, which is currently serving up to four billion hosts over diverse networks. Despite this, IPv4 has still been successfully functioning well since 1981. Over the last couple of years, the massive growth of the Internet has been evident requiring an evolution of the whole architecture of the Internet Protocol. Therefore, in order to strengthen the existing architecture of Internet Protocol, IETF has developed Internet Protocol version 6 (IPv6) [1]. IPv6 offers a significant improvement over IPv4 when it comes to the unlimited address space, the built-in mobility and the security support, easy configuration of end systems, and enhanced multicast features, etc [2]. On the other hand, due to the fascination of end users of the World Wide Web (WWW) and the popularity of real-time applications, we can now observe new increasing demands on real-time multimedia services over the Internet. As the name implies, these services have timing constraints due to their real-time nature. For instance, video and voice applications typically have bandwidth, delay and loss requirements when the data does not arrive in time turning the play out process paused, which is annoying to the end users [3].

In such a new environment, as the expansion of the Internet continues, QoS is a basic requirement in terms of provisioning the multimedia services where deployment of IPv6 QoS is underway. Today, there is limited experience in the deployment of QoS for IPv6 traffic in MPLS backbone networks in conjunction with DiffServ support. Many organizations and groups are still working in order to ensure a guaranteed service for the real-time applications as a framework to the Internet. In that case, the IETF has introduced several service models, mechanisms, policies and schemes for satisfying QoS demands. DiffServ [4] and MPLS [5] are known as notable mechanisms to provide QoS guarantee [6]. The DiffServ architecture model provides the most extended and attractive solution for QoS support in IPv4/IPv6 networks. Scalability and traffic classification are main concerns for DiffServ as it can handle large number of data networks very efficiently.

This is accomplished through the combination of traffic conditioning and Per-Hop Behaviour based (PHB) forwarding by using the field DSCP (Differentiated Service Code Point) [7]. This field exists in both IPv4 and IPv6 packet headers. MPLS is a network protocol technology that helps to improve scalability and routing flexibility in IP networks. The conventional IP network creates hot spots (Hyper-aggregation) on the shortest distance path between two points while other alternative paths remain underutilized. For this circumstance, IP network can experience some problems such as longer delay, degradation in the throughput and packet losses. In such a situation, MPLS TE is best suited for minimizing the effects of congestion by bandwidth optimization [8].

Combination of two QoS mechanisms (DiffServ and MPLS) has already been evaluated and experimented on IPv4 environment whereas the deployment of IPv6 in MPLS networks, one of the approaches called IPv6 MPLS over IPv4-based core (6PE) has been undertaken in a greater extent. In terms of IPv6 deployment in MPLS networks, there are four approaches including IPv6 over a circuit transport over MPLS, IPv6 over IPv4 tunnels over MPLS, IPv6 MPLS with IPv4-based core (6PE), and IPv6 MPLS with IPv6-based core [9]. In such a case performance evaluation of IPv6 MPLS with IPv6-based core in conjunction with DiffServ has not yet been elaborately evaluated and experimented so far. Therefore, in this paper, a comparative study has been done on the performance evaluation of video and voice over DiffServ-MPLS in IPv4/IPv6 networks. The research question to be examined in this paper is formulated as follows: to what extent does the performance of PDV for AF and EF PHBs vary from DiffServ-MPLS/IPv4 network to DiffServ-MPLS/IPv6 network?

## 2. RELATED WORK

Several researches have concentrated on the IPv6 network performance over the last decade. There is a very large literature on general aspects of the new Internet Protocol IPv6 and its QoS evaluation are described in [10] [11] [12] [13] [14] [15] [16] [17] [18].

The authors in [10] have described the implementation of a testbed and the inter–connection between three DiffServ domains by using IPv6–in–IPv4 static tunnels. They have investigated the performance issues like throughput, packet loss and delay of particularly aviation applications such as Controller to Pilot Data Link Communication by using DiffServ on the IPv6-based backbone network. Their obtained results confirm that the DiffServ implementation and support in IPv6 network has been matured enough to provide stable and reliable QoS for the aviation applications.

In [11], authors have evaluated the performance of TCP and UDP transmission under different network environment e.g., a pure IPv4, a pure IPv6 and an IPv6 tunnelled in IPv4 using MPLS Linux environments. With regard to the environment construction and measurement tool they have used virtual machine running on Linux and Iperf, respectively. They have showed that the performance of TCP transmissions in both IPv4 and IPv6 is almost the same while the performance of TCP transmission in tunnelling of IPv6 in IPv4 using MPLS Linux is lower than the pure IPv4 and IPv6 performance while the performance of UDP transmission in all three different environments is close to each other.

In [12] authors have presented whether a QoS service implemented on a large-scale native IPv6 network works well. In their investigation, they have concluded that the QoS mechanisms (i.e., classification, prioritization, policing) perform well.

The authors in [15] have presented on how to deploy DiffServ in order to assess priority functionalities. In their paper, a scheduling mechanism based on WRED for the Intel® IXP2400 network processor has been developed and tested to provide QoS by maintaining priority of incoming packets based on criteria i.e. class of packets and traffic.

The authors in [16] have done a comparison between three QoS schemes i.e., Integrated Service, DiffServ and IPv6 QoS Management with respect to QoS guarantee. In comparison of their achieved results from the test show that IPv6 QoS management scheme achieves the best results during conformant and non-conformant test compared to both IntServ and DiffServ schemes.

From the above described related work, it is observed that most of the works have been done so far followed by the method, experimental measurement. Furthermore, none of the above research work has done a simulative evaluation of real-time applications such as video and voice performance in terms of PDV in relation to DiffServ with MPLS TE in the IPv4/IPv6 networks. In this work, realizing a simulation approach, a comparative performance analysis of video and voice conferencing in conjunction with DiffServ with or without MPLS TE has been complemented.

## 3. BACKGROUND

### 3.1 IPv6 Implementation over MPLS network

Several approaches are possible to offer IPv6 connectivity over the MPLS core domain. They vary from a couple of standpoints: transitioning strategy, scalability, data overhead, and configuration. IPv6 MPLS with IPv6-based core compares the different solutions in relation to the support of IPv6 in MPLS [9].

### 3.2 Packet Delay Variation (PDV)

The performance metric, PDV is based on the difference in the One-Way-Delay (OWD) of selected packets. This difference in delay is called "IP Packet Delay Variation (IPDV)" as defined in a draft of the IETF IPPM working group [19].

## 4. NETWORK MODEL AND IMPLEMENTATION

### 4.1 Network Traffic Generation

Detailed information about the configurable parameters for voice applications is given in Table 1, 2 and 3. In voice applications, voice traffic configuration we have set the codec bit rate at 64 Kbps and codec sample interval 10 ms whereby codec sample size is calculated using 64,000 *10/1000 = 640 bits (e.g., codec bit rate=sample interval/sample size). Thus the sample size is 80 bytes. For 10 ms sample interval 100 packets per second needs to be transmitted [20].

Video and voice conferencing profiles are defined in the source workstations while corresponding destination workstations are enabled with their respective supported services. In OPNET terminology, in order to generate voice and video traffic, voice and video conferencing profiles are configured in such a way where video and voice applications can be controlled in terms of their start, end times and repeatability. This is done by adding this profile to each workstation's lists of supported profiles. The start time and offset time for the video_and_voice_profile configuration parameters are presented in Table 1. It is noted that while configuring the profile for video and voice conferencing; the first call by each designated workstation starts at 120 seconds (i.e. start time of 100 seconds with offset time of 20 seconds), while the second call is added at 420 seconds of simulation time, and finally the third call is added at 720 seconds of the simulation time (1800 seconds). Which follows each designated workstation is having three interactive video conferencing sessions running simultaneously during the simulation period (i.e. 720-1800 seconds).

Table 1. Voice and video profile configuration parameters

| Profile Name | Applications | Operation Mode | Start Time (seconds) | Duration (seconds) | Repeatability |
|---|---|---|---|---|---|
| Video Traffic_AF11 Profile | (...) | Simultaneous | constant (100) | End of Simulation | Once at Start Time |
| Video Traffic_AF12 Profile | (...) | Simultaneous | constant (100) | End of Simulation | Once at Start Time |
| Video Traffic_AF13 Profile | (...) | Simultaneous | constant (100) | End of Simulation | Once at Start Time |
| Video Traffic_AF41 Profile | (...) | Simultaneous | constant (100) | End of Simulation | Once at Start Time |
| Video Traffic_AF42 Profile | (...) | Simultaneous | constant (100) | End of Simulation | Once at Start Time |
| Video Traffic_AF43 Profile | (...) | Simultaneous | constant (100) | End of Simulation | Once at Start Time |
| Voice PCM Quality_EF Profile | (...) | Simultaneous | constant (100) | End of Simulation | Once at Start Time |

| Name | Start Time Offset (seconds) | Duration (seconds) | Repeatability |
|---|---|---|---|
| Video Conference AF11_10Frame | constant (20) | End of Profile | (...) |
| Video Conference AF11_15Frame | constant (320) | End of Profile | (...) |
| Video Conference AF11_30Frame | constant (620) | End of Profile | (...) |

Table 2. Voice and video application configuration parameters

| Video_and_Voice_Profiles | Frame Size [Bytes] | Bit Rate [Kbps] | Total Offered Load[Kbps] | Start-time [s] |
|---|---|---|---|---|
| Video Conference _AF11_10Frame | 4000 | 320 |  | 120 |
| Video Conference _AF11_15Frame | 4000 | 480 | 1760 | 420 |
| Video Conference _AF11_30Frame | 4000 | 960 |  | 720 |
| Video Conference _AF12_10Frame | 3000 | 240 |  | 120 |
| Video Conference _AF12_15Frame | 3000 | 360 | 1320 | 420 |
| Video Conference _AF12_30Frame | 3000 | 720 |  | 720 |
| Video Conference _AF13_10Frame | 2000 | 160 |  | 120 |
| Video Conference _AF13_15Frame | 2000 | 240 | 880 | 420 |
| Video Conference_AF13_30Frame | 2000 | 480 |  | 720 |
| Video Conference_AF41_10Frame | 3500 | 280 |  | 120 |
| Video Conference_AF41_15Frame | 3500 | 420 | 1540 | 420 |
| Video Conference_AF41_30Frame | 3500 | 840 |  | 720 |
| Video Conference_AF42_10Frame | 2500 | 200 |  | 120 |
| Video Conference_AF42_15Frame | 2500 | 300 | 1100 | 420 |
| Video Conference_AF42_30Frame | 2500 | 600 |  | 720 |
| Video Conference_AF43_10Frame | 1500 | 120 |  | 120 |
| Video Conference_AF43_15Frame | 1500 | 180 | 760 | 420 |
| Video Conference_AF43_30Frame | 1500 | 360 |  | 720 |
| Voice PCM Quality_EF | 80 | 64 |  | 120 |
| Voice PCM Quality_EF | 80 | 64 | 270 | 420 |
| Voice PCM Quality_EF | 80 | 64 |  | 720 |

Table 3. Voice application parameters

| Attribute | Value | |
|---|---|---|
| Silence Length (sec) | Incoming Silence Length (sec) | Exponential (0.65) |
|  | Outgoing silence Length (sec) | Exponential (0.65) |
| Encoder scheme | G.711 | |
| Voice Frames per packet | 1 | |
| Type of Service | Best Effort (0) | |
| Compression Delay (sec) | 0.02 | |
| Decompression Delay (sec) | 0.02 | |

**4.2 Simulation Scenarios**

OPNET Modeler 14.0 [27] has been used for the simulation analysis. This section explains the network model used in this study. Six network scenarios have been prototyped as follows, which will be elaborately demonstrated in the upcoming sections. Scenario 1 is modelled as a baseline scenario without QoS implementation. Scenario 2 serves as another baseline scenario to demonstrate traffic delivery in a best-effort IPv6 network under congested condition in which no QoS is configured. Scenario 3 is modelled followed by baseline scenario 1 where DiffServ has been implemented while scenario 4 is modelled followed by baseline scenario 2 with DiffServ implementation. Scenario 5 is modelled to demonstrate real-time applications delivery in a DiffServ enabled MPLS network followed by scenarios 1 and 3 while scenario 6 modelled followed by scenarios 2 and 4. It is important to mention that in all the simulation scenarios, the routers ethernet2_slip8_ler and ethernet2_slip8_lsr [27] correspond to the LERs and LSRs, respectively. These routers are interconnected via ppp_adv point-to-point link operated at 4Mbps data rate. The links used to connect switches with the routers (i.e. LER1 and LER2) are 100Base-T, while 10Base-T is to connect the workstations with the switches. The switches namely switch_1 and switch_2 (i.e. ethernet16_switch) are connected with routers ((i.e. ethernet2_slip8_ler and ethernet2_slip8_lsr)) using 100Base-T. The scenarios to be modelled in this work are outlined as follows:

- Scenario 1: Baseline IPv4 Network

- Scenario 2: Baseline IPv6 Network

- Scenario 3: DiffServ without MPLS TE in IPv4 Network

- Scenario 4: DiffServ without MPLS TE in IPv6 Network

- Scenario 5: DiffServ with MPLS TE in IPv4 Network

- Scenario 6: DiffServ with MPLS TE in IPv6 Network

### 4.1.1 Scenario 1: Baseline_IPv4

In order to study the results from other scenarios (3, 4, 5 and 6), a baseline network model considering a typical meshed IP network has been prototyped in which packets are forwarded from IPv4 source to the corresponding IPv4 destination through the IPv4 core domain with the best-effort policies.

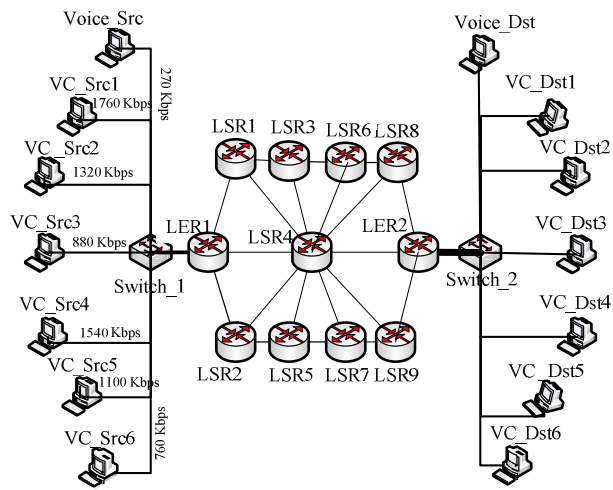

Figure 1. Baseline IPv4 Network Topology

In this scenario, each pair uses a best-effort service as a Type of Service (ToS). All routers (LERs and LSRs) in the given baseline topology are DiffServ and MPLS TE disabled. The reference network topology depicted in Figure 1 is composed of six pairs of video conferencing workstations and a pair of voice workstations. The core network consists of nine LSRs (i.e. Label Switched Router) and two LERs (i.e. Label Edge Router). All the LSRs and LERs of the core network are interconnected using the point-to-point link (ppp_adv) operated at a 4Mbps

32

data rate. In our reference network topology (Figure 1), OSPF [21] routing protocol is used under normal condition without considering load balancing feature and MPLS is set to disable. The purpose of not considering load balancing is that MPLS TE can be better understood.

**4.1.2 Scenario 2: Baseline_IPv6**

Topology depicted in Figure 1 represents scenario 2 which is same as the scenario 1, but IPv6 is configured in scenario 2. All IP nodes in the scenario 1 are dual-stack capable supporting both IPv4 and IPv6. In this scenario, to manually configure an interface to support IPv6 only but not IPv4, the IPv4 address of the interface is set to *No IP Address*. IPv6 link-local and global addresses on interfaces of all nodes in the network have manually been configured. In order to configure IPv6 in the network, Link-Local Address attribute is set to Default EUI-64 while Global Address (es) is set to *EUI-64* with the specification of the first 64 bits of the address. The remaining 64 bits of the address are set to an interface ID unique to the interface. With regard to routing protocol configuration of IPv6 network, as the process v2 of OSPFv2 is already running for IPv4 network (scenario 1). In this scenario, the process v2 has been disabled while another process version (v3) [22] is enabled to the OSPF parameters configuration.

**4.1.3 Scenario 3: DiffServ without MPLS_IPv4**

In order to configure scenario 3, the following configuration is made with scenario 1. The main goal of this scenario lies in the differentiation of flows at an edge router (LER1) of a DS-domain. Abstraction of DiffServ QoS configuration involved in this scenario is described in the following section:

- Traffic classification and marking

- Scheduling

- Configuring Class-Based DSCP WRED

- Traffic Policy Configuration

**4.1.3.1 Traffic Classification and Marking**

DiffServ QoS relies on the classification of traffic, to provide different quality-of-service level on a per-hop basis. Traffic can be classified based on a wide variety of criteria called traffic descriptors, which include: ToS value in an IP header (IP Precedence or DSCP). Configuration of Extended Access Lists (ACLs) presented in Table 4 is used to identify video and voice traffics for classification based on source address of workstations. After classification, traffic should be marked to indicate the required level of QoS service for that traffic. Marking can occur within either the Layer-2 header or the Layer-3 header [23]. In layer 3 marking, there are two marking methods where one uses the first three bits of the ToS field and other one uses the first six bits of the ToS field (DSCP) [24]. In our test case, layer-3 marking for voice and video traffic is accomplished based on DSCP where the traffic has been marked on the inbound interface of edge router, LER1. Now marked traffic flows are subjected to forwarding behavior based on their corresponding DSCP value. This forwarding behavior is implemented using CBWFQ and WRED. There are two standard PHB groups: Assured Forwarding (AF) PHB and Expedited Forwarding (EF) PHB where an AF PHB group consists of four AF classes; AF1x through AF4x [25]. Looking at the network topology exhibited in Figure 1, video conferencing traffic flows generated by source workstations e.g., VC_Src1, VC_Src2, VC_Src3, VC_Src4, VC_Src5, and VC_Src6 are marked with AF11, AF12, AF13, AF41, AF42, and AF43 classes, respectively. The EF PHB is used for voice traffic flows as it provides a low loss, low latency, low jitter, assured bandwidth, and end-to-end services.

Table 4. IPv4 extended ACL configuration

| ACL Name | Action | Source | | | DSCP |
|---|---|---|---|---|---|
| | | Workstations | IP | Wildcard Mask | |
| EF | Permit | Voice_Src | 192.0.17.1 | 0.0.0.255 | EF |
| AF11 | Permit | Vc_Src1 | 192.0.17.2 | 0.0.0.255 | AF11 |
| AF12 | Permit | Vc_Src2 | 192.0.17.3 | 0.0.0.255 | AF12 |
| AF13 | Permit | Vc_Src3 | 192.0.17.4 | 0.0.0.255 | AF13 |
| AF41 | Permit | Vc_Src4 | 192.0.17.5 | 0.0.0.255 | AF41 |
| AF42 | Permit | Vc_Src5 | 192.0.17.6 | 0.0.0.255 | AF42 |
| AF43 | Permit | Vc_Src6 | 192.0.17.7 | 0.0.0.255 | AF43 |

**4.1.3.2 Scheduling**

Seven CBWFQ profiles (see detail in Table 5) are defined under IP QoS Parameters in OPNET. The amount of bandwidth in percentage of available bandwidth is assigned to the seven traffic classes. In our case, bandwidth type is assigned to *Relative* that means if a traffic class does not use or need bandwidth equal to the reserved, available bandwidth can be used by other bandwidth classes. Queue limit is set to 500 Packets while priority is set to *Enable* with EF class. EF class carries voice traffic which is delay and loss sensitive. Setting the priority as *Enable* provides strict priority for CBWFQ and allows voice traffic to be dequeued and sent before packets in other queues are dequeued.

**4.1.3.3 Configuring Class-Based DSCP WRED**

WRED is an extension to RED. It allows configuring different drop profiles to different traffic flows and providing different QoS for different types of traffic.

Table 5. Typical DiffServ queue bandwidth allocation (CBWFQ Profiles).

| CBWFQ Profile Name | BW. Type | BW (%) | Queue Limit Pckt. | WRED Profiles. | | | | |
|---|---|---|---|---|---|---|---|---|
| | | | | Match Property | Exp. Wei Cons. | Min. Th. Pckt. | Max. Th. Pckt. | Mark Prob. Dnmtr. |
| WFQ_EF | Relative | 5 | 500 | DSCP | 9 | 100 | 200 | 10 |
| WFQ_AF11 | Relative | 20 | 500 | DSCP | 9 | 100 | 200 | 10 |
| WFQ_AF12 | Relative | 10 | 500 | DSCP | 9 | 100 | 200 | 10 |
| WFQ_AF13 | Relative | 5 | 500 | DSCP | 9 | 100 | 200 | 10 |
| WFQ_AF41 | Relative | 40 | 500 | DSCP | 9 | 100 | 200 | 10 |
| WFQ_AF42 | Relative | 15 | 500 | DSCP | 9 | 100 | 200 | 10 |
| WFQ_AF43 | Relative | 5 | 500 | DSCP | 9 | 100 | 200 | 10 |

Seven WRED profiles defined under *IP QoS Parameters* attribute which is able to distinguish traffic flows by examining DSCP value. Detail configuration parameters of WRED profiles are provided in Table 5. It is noted that each QoS attribute is configured at the output interfaces of the edge router (LER1).

International Journal of Distributed and Parallel Systems (IJDPS) Vol.3, No.1, January 2012

**4.1.3.4 Traffic Policy Configuration**

In OPNET modeler suite, traffic policies can be defined and configured on the inbound/outbound interface of routers under the IP QoS *Parameters Traffic Policies* attribute. Table 6 illustrates QoS mechanisms such as scheduling (CBWFQ) and policing (WRED) are grouped into the defined traffic policy (Traffic_Policy) and applied to corresponding traffic classes where each Traffic Class referenced in the traffic policy is associated with two profiles. An outbound traffic policy, Traffic_Policy is applied to the outbound interface of the edge router, LER1 as scheduling and congestion avoidance are supported only in the outbound direction.

Table 6. Class-Based WFQ (CBWFQ) profiles

| Policy Name | Configuration | | |
|---|---|---|---|
| | Traffic Class Name | Set Info | |
| | | Set Property | Set Value |
| Traffic_Policy | EF | WFQ Profile (Class Based) | WFQ_ EF |
| | | RED/WRED Profile | WRED_EF |
| | AF11 | WFQ Profile (Class Based) | WFQ_ AF11 |
| | | RED/WRED Profile | WRED_AF11 |
| | AF12 | WFQ Profile (Class Based) | WFQ_ AF12 |
| | | RED/WRED Profile | WRED_AF12 |
| | AF13 | WFQ Profile (Class Based) | WFQ_ AF13 |
| | | RED/WRED Profile | WRED_AF13 |
| | AF41 | WFQ Profile (Class Based) | WFQ_ AF41 |
| | | RED/WRED Profile | WRED_AF41 |
| | AF42 | WFQ Profile (Class Based) | WFQ_ AF42 |
| | | RED/WRED Profile | WRED_AF42 |
| | AF43 | WFQ Profile (Class Based) | WFQ_ AF43 |
| | | RED/WRED Profile | WRED_AF43 |

**4.1.4 Scenario 4: DiffServ without MPLS_IPv6**

This scenario is configured based on the scenarios 2 and 3. Additional configuration involved in implementing IPv6 QoS is discussed below. For IPv6 QoS implementation in the network topology depicted in Figure 1, all relevant factors including network equipments and application in the network are capable to support IPv6 QoS. IPv6 Header has two segments relevant with QoS, TC (Traffic Class) and FL (Flow Label) [7]. TC has 8 bits and same as the ToS in IPv4. The Traffic Class field is used to set DSCP values. These values are used in the exact same way as in IPv4. In this scenario, classification is accomplished based on IPv6 precedence, DSCP which is defined in the configured extended IPv6 access lists. After traffic classification, in order to carry out IPv6 DiffServ implementation, the steps needed to be followed are described in scenario 3.

**4.1.5 Scenario 5: DiffServ with MPLS_IPv4**

The DiffServ configuration for MPLS network models is similar to IP QoS configuration in DiffServ IPv4 network that is not MPLS-enabled. The main difference for MPLS networks is that packet is marked with the appropriate EXP bits according to their traffic class at edge routers, LER1 and LER2. The goal of this scenario is to minimize congestion by making some traffic follow the "non-shortest path" through the network and distributing the total real-time traffic across the pre-established LSPs according to the current state of the network. The example topology presented in Figure 1 is considered followed by the network configuration of scenarios 1 and 3.

For deploying MPLS TE in the network, it is important to determine whether or not TE is required for a given network. This determination has been made by doing an IGP analysis on the scenario 3. The IGP analysis is done by running a Discrete Event Simulation (DES) of the scenario 3, which shows that one link, *LER1→LSR4* is over-utilized while other links *LER1→LSR2* and *LER2→LSR1* are unused that turns out to have a need of TE configuration and analysis. In this case, in terms of TE implementation and analysis of the network, the next step is to create LSPs in the network that will direct traffic from the over-utilized links towards the less utilized links. In the OPNET MPLS model suite, *Global MPLS Attributes* are used to configure network-wide MPLS parameters that are grouped in the MPLS configuration object, MPLS_Config. Router specific MPLS attributes are grouped in the *MPLS Parameters Attribute* on each core router of DiffServ/MPLS IPv4 domain.

### 4.1.5.1 MPLS TE Configuration in the Network

This section describes how to manually configure MPLS TE in IPv4 network using OPNET MPLS Model Suite. The following topics are covered by configuring LSPs and defining how traffic is assigned to the corresponding LSPs. Before LSPs are configured, status of MPLS on the Interfaces running OSPF of core routers of DiffServ/IPv4 domain is set to *Enable*. The edge routers, LER1 and LER2 are considered as the source and destination of the LSPs, respectively. In order to make LSPs reachable from other sections of the MPLS domain, a loopback interface on the routers has been configured. Configuring MPLS in a network can be split in a three-step process as follows.

### 4.1.5.2 LSPs Creation and Configuration in the Network Topology

Static LSPs are created using the path object, MPLS_E-LSP_STATIC. In our proposed network, six bidirectional LSPs are created namely LSP1_0, LSP1_1, LSP1_2, LSP2_0, LSP2_1 and LSP2_2 in a way that they can be initiated on both LER1 and LER2. Motivation of using static LSPs is that it allows more routing control but it has fewer resiliencies to link failures, however, link failures is out of scope of this paper. One of the important LSP attributes is that for E-LSP, three experimental bits in the shim header carry the DiffServ information. This provides eight different ToS per LSP [26].

### 4.1.5.3 FECs and Traffic Trunks Creation and Configuration in the MPLS_Config

Table 7 presents traffic trunk's profiles that are aggregates of traffic flows belonging to the same or different classes. Forwarding Equivalence Class (FEC) parameters are used to classify and group packets, so that all packets in a group are forwarded the same way. In order to do that seven FECs are defined based on DSCP in MPLS_Config. Each of FECs consists of three UDP traffic flows that are treated as traffic aggregate in the MPLS domain. For example, *FEC for AF11* is identified by this name when TE assignments are specified in *Traffic Mapping Configuration*, defining the criteria for the FECs in it. Seven traffic trunk profiles are created based on seven DiffServ codes in the MPLS_Config, which specifies out-of-profile actions and traffic classes for traffic trunks in the network. Traffic trunks capture traffic characteristics such as peak rate, average rate, and average burst size. The detail out-of-profile settings of the traffic trunk profiles can be found in Table 7.

Table 7. Traffic (Trunk Profile)

| Trunk Name | Trunk Details | | Value |
|---|---|---|---|
| Trunk for Video Traffic AF11 | Traffic Profile | Maximum Bit Rate(bits/s) | 2,000,000 |
| | | Peak Burst Size (bits) | 2,000,000 |
| | | Average Bit Rate (bits/s) | 2,000,000 |
| | | Maximum Burst size(bits) | 2,000,000 |
| | Out of profile | Out of profile Action | Transmit |
| | | Remark Precedence | Transmit Unchanged |
| | Traffic Class | AF11 | |
| Trunk for Video Traffic AF12 | Traffic Profile | Maximum Bit Rate(bits/s) | 2,000,000 |
| | | Peak Burst Size (bits) | 2,000,000 |
| | | Average Bit Rate (bits/s) | 2,000,000 |
| | | Maximum Burst size(bits) | 2,000,000 |
| | Out of profile | Out of profile Action | Transmit |
| | | Remark Precedence | Transmit Unchanged |
| | Traffic Class | AF12 | |
| Trunk for Video Traffic AF13 | Traffic Profile | Maximum Bit Rate(bits/s) | 2,000,000 |
| | | Peak Burst Size (bits) | 2,000,000 |
| | | Average Bit Rate (bits/s) | 2,000,000 |
| | | Maximum Burst size(bits) | 2,000,000 |
| | Out of profile | Out of profile Action | Transmit |
| | | Remark Precedence | Transmit Unchanged |
| | Traffic Class | AF13 | |
| Trunk for Video Traffic AF41 | Traffic Profile | Maximum Bit Rate(bits/s) | 3,000,000 |
| | | Peak Burst Size (bits) | 3,000,000 |
| | | Average Bit Rate (bits/s) | 3,000,000 |
| | | Maximum Burst size(bits) | 3,000,000 |
| | Out of profile | Out of profile Action | Transmit |
| | | Remark Precedence | Transmit Unchanged |
| | Traffic Class | AF41 | |
| Trunk for Video Traffic AF42 | Traffic Profile | Maximum Bit Rate(bits/s) | 3,000,000 |
| | | Peak Burst Size (bits) | 3,000,000 |
| | | Average Bit Rate (bits/s) | 3,000,000 |
| | | Maximum Burst size(bits) | 3,000,000 |
| | Out of profile | Out of profile Action | Transmit |
| | | Remark Precedence | Transmit Unchanged |
| | Traffic Class | AF42 | |
| Trunk for Video Traffic AF43 | Traffic Profile | Maximum Bit Rate(bits/s) | 3,000,000 |
| | | Peak Burst Size (bits) | 3,000,000 |
| | | Average Bit Rate (bits/s) | 3,000,000 |
| | | Maximum Burst size(bits) | 3,000,000 |
| | Out of profile | Out of profile Action | Transmit |
| | | Remark Precedence | Transmit Unchanged |
| | Traffic Class | AF43 | |

**4.1.5.4 Configuring LERs to Direct Packets into the Appropriate LSPs**

Some of the important MPLS parameters are set on the edge routers, LER1 and LER2 which are described in this section. Traffic Mapping Configuration specifies bindings between FECs and LSPs. In Table 8, each row of the traffic mapping configuration specifies a distinct TE binding in which each TE binding specifies FEC, traffic trunk, and LSP that is applied to the label of the incoming packet. For instance, a FEC such as *FEC for AF11* is bound to a traffic trunk, *Trunk for Video Traffic AF11* which is mapped on to LSP1_0. These mappings defined in the MPLS_Config are used by the edge routers. A standard *EXP<=>PHB* mapping is applied to determine the PHB of the behavior aggregates that are mapped onto a single E-LSP.

Table 8. Traffic mapping configuration

| Interface In | FEC/Destination Prefix | DSCP | Traffic Trunk Profiles | LSP |
|---|---|---|---|---|
| 8 | FEC For AF11 | AF11 | Trunk_for_Video_Traffic_AF11 | LSP1_0 |
| 8 | FEC For AF43 | AF12 | Trunk_for_Video_Traffic_AF43 | LSP1_0 |
| 8 | FEC For AF13 | AF13 | Trunk for Video Traffic AF13 | LSP1_1 |
| 8 | FEC For AF42 | AF41 | Trunk for Video Traffic AF42 | LSP1_1 |
| 8 | FEC For AF12 | AF42 | Trunk for Video Traffic AF12 | LSP1_2 |
| 8 | FEC For AF41 | AF43 | Trunk for Video Traffic AF41 | LSP1_2 |
| 8 | FEC For Voice EF | EF | Trunk For Voice Traffic | LSP1_2 |

### 4.1.6  Scenario 6: DiffServ with MPLS_IPv6

Scenario 6 is similar to scenario 4 except the deployment of IPv6 support on MPLS network. Several approaches are possible to offer IPv6 connectivity over the MPLS core domain. They vary from a couple of standpoints: transitioning strategy, scalability, data overhead, and configuration in relation to the support of IPv6 in MPLS [9]. In our case of IPv6 support on MPLS network, approach 4, *IPv6 MPLS with IPv6-based Core* is considered where all the LSRs are configured in such a way that it can support IPv6 completely.

## 5. RESULTS ANALYSIS

### 5.1 PDV Performance for AF11 Traffic Flows

Figure 2 shows the comparable performance of the PDV under the different network scenarios measured for the video conferencing traffic flows, which are generated at VC_Src1(source node) and destined to VC_Dst1 (destination node) through the IPv4 and IPv6 networks. The comparison against different network scenarios in terms of PDV is referred to Table 9, and observed in Figure 2.

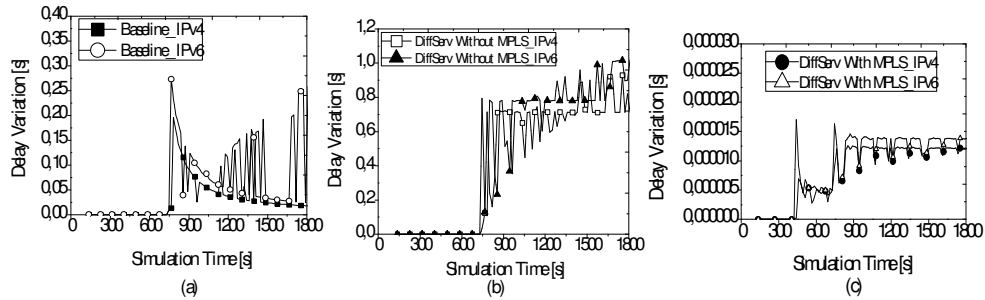

Figure 2. PDV experienced by AF11 flows for scenarios (a) 1-2 (b) 3-4 and (c) 5-6.

In the case of baseline networks, from Figure 2(a), one can examine that PDV in IPv4 varies from 0.1 ns (nanosecond) to 196 ms (millisecond) along with an average value 35 ms, whereas PDV in IPv6 varies from 0.3 ns to 273 ms followed by an average value 64 ms.

In the case of DiffServ IPv4/IPv6 networks exhibited in Figure 2(b), and Table 9 follows that PDV for AF11 traffic flows in IPv4 is differed from 0.0 s to 930 ms achieving an average 427 ms while PDV in IPv6 is differed from 0.0 ms to 1019 ms with an average 490 ms. From DiffServ perspective, the average PDV of AF11 in both IPv4 and IPv6 network is considerably higher than scenarios 1 and 2. This is caused by as weight set to 20% of the link capacity (4Mbps) and queue size set at 500 packets. This can be explained in way that VC_Src1 is generating total video traffic about 1.7 Mbps. In order to successfully transmit this traffic, the required bandwidth is about 1.9 Mbps including additional overhead by layer protocols. But the assigned weight is about 0.8 Mbps. As a result, it exceeds the link capacity. In summary, from the IPv6 protocol performance perspective, PDV in DiffServ IPv6 network is fairly about 12% higher than that of counterpart, IPv4.

In the case of scenarios 5 and 6 depicted in Figure 2 (c), and from Table 9, it is observed that PDV for AF11 traffic flows in DiffServ/MPLS IPv4 network differs from approximately 0.0 s to 17 $\mu$s (microsecond), attaining an average PDV of about 8 $\mu$s, alternatively PDV for AF11 in DiffServ/MPLS IPv6 network varies from 0.1 ns to about 26 $\mu$s with an average 9 µs. Table 9 depicts the average PDV which is very small compare to the scenarios 3 and 4. That means adopting MPLS TE in DiffServ network improves PDV performance for video conferencing traffic. In the context of the IPv4/IPv6 protocol performance, AF11 in DiffServ/MPLS IPv6 network still suffers 7% higher than that of IPv4.

Based on the simulation results illustrated in Table 9 and Figure 2, it can be concluded that on an average, PDV for AF11 in DiffServ IPv6 network is appeared to be 12% higher than that of counterpart IPv4 while PDV for AF11 in DiffServ/MPLS TE IPv6 network remains 7% higher than IPv4.

Table 9. Summary statistics of PDV experienced by AF11 flows.

| Scenarios | Min. [s] | Avg. [s] | Max. [s] | Std Dev [s] |
|---|---|---|---|---|
| Scenario 1 | 1,00E-10 | 3,52E-02 | 1,96E-01 | 4,61E-02 |
| Scenario 2 | 3,00E-10 | 6,46E-02 | 2,73E-01 | 7,64E-02 |
| Scenario 3 | 0,00E+00 | 4,28E-01 | 9,30E-01 | 3,53E-01 |
| Scenario 4 | 1,00E-10 | 4,90E-01 | 1,02E+00 | 4,00E-01 |
| Scenario 5 | 0,00E+00 | 8,48E-06 | 1,70E-05 | 4,79E-06 |
| Scenario 6 | 1,00E-10 | 9,13E-06 | 2,63E-05 | 5,58E-06 |

## 5.2 PDV Performance for AF12 Traffic Flows

Figure 3 illustrates the comparable performance of the PDV under the different network scenarios measured for the video conferencing traffic flows generated at VC_Src2 and destined to VC_Dst2 through the IPv4 and IPv6 networks. The comparison against different network scenarios in terms of PDV is referred to Table 10, and observed in Figure 3.

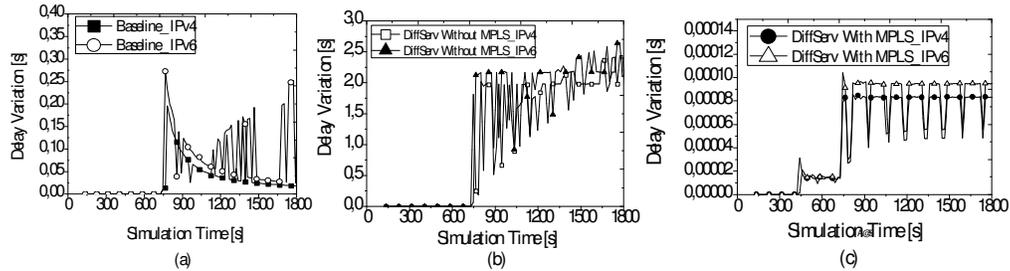

Figure 3. PDV experienced by AF12 flows for scenarios (a) 1-2 (b) 3-4 and (c) 5-6.

In support of the DiffServ in the baseline network scenarios shown in Figure 3(b), where Table 10 follows that PDV for the AF12 traffic flows in IPv4 is differed from 1.5 ns to 2.43 s achieving an average 1.15 s. At the same time, one can visualize in IPv6, PDV for AF12 varies from 1 ns to 2.6 s with an average 1.27 ms. From the IPv6 protocol performance perspective, PDV in DiffServ IPv6 network is relatively 10% higher than that of IPv4. From DiffServ perspective, in both scenarios 3 and 4, the offered traffic load by VC_Src1 is 50% higher than the assigned weight 10% (0.4 Mbps) for AF12 with medium priority which leads to the higher delay compare to the base-line scenarios 1 and 2.

In the case of scenarios 5 and 6 shown in Figure 3 (c), Table 10 indicates that the PDV for AF12 traffic flows in DiffServ/MPLS IPv4 network differs from approximately 0.0 $\mu$s to 85 $\mu$s, achieving an average about 50 $\mu$s while PDV in IPv6 is varied from 0.0 $\mu$s to roughly 103 $\mu$s with an average 57 $\mu$s. By examining the obtained results of PDV with regard to the IPv6 protocol performance, AF12 in the IPv6 network perceives 11% higher PDV than that of IPv4 when TE is considered in the IP/DiffServ. In addition, from the Figure 3 and Table 10, it can be concluded that on an average, in the case of IP/DiffServ network, PDV for AF12 IPv6 contributes to 10% higher PDV than counterpart IPv4, while PDV for AF12 in the DiffServ/MPLS TE IPv6 network remains 11% higher than IPv4.

Table 10. Summary statistics of PDV experienced by AF12 flows

| Scenarios  | Min. [s]   | Avg. [s]  | Max. [s]  | Std Dev [s] |
|------------|------------|-----------|-----------|-------------|
| Scenario 1 | 1,00E-10   | 3,52E-02  | 1,96E-01  | 4,61E-02    |
| Scenario 2 | 3,00E-10   | 6,46E-02  | 2,73E-01  | 7,64E-02    |
| Scenario 3 | 1,50E-09   | 1,15E+00  | 2,44E+00  | 9,52E-01    |
| Scenario 4 | 1,00E-10   | 1,28E+00  | 2,65E+00  | 1,04E+00    |
| Scenario 5 | 0,00E+00   | 5,08E-05  | 8,54E-05  | 3,53E-05    |
| Scenario 6 | 0,00E+00   | 5,71E-05  | 1,04E-04  | 4,08E-05    |

## 5.3 PDV Performance for AF13 Traffic Flows

Figure 4 illustrates the comparable performance of the packet delay variation (PDV) under the different network scenarios measured for the video conferencing traffic flows generated at VC_Src3 and destined to VC_Dst3 through the IPv4 and IPv6 networks. The comparison against different network scenarios in terms of PDV is referred to Table 11, and observed in Figure 4.

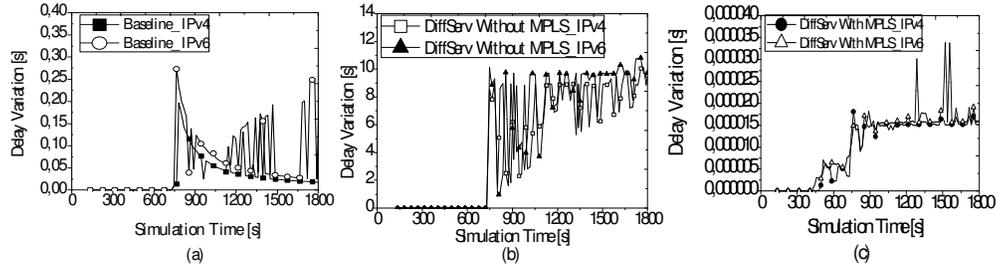

Figure 4. PDV experienced by AF13 flows for scenarios (a) 1-2 (b) 3-4 and (c) 5-6

In scenarios 3 and 4, one can see from Table 11 that PDV for AF13 traffic flows in IPv4 network exhibited in Figure 4(b) is differed from 0.0 ns to 10.2s with the average value 4.7 s. At the same time, PDV experienced in IPv6 network by AF13 traffic flows (Figure 4(b)) varies from 1 ns (e.g. 40% network load) to 10.8 s (200% network load) with an average 5.1 s. PDV introduced in DiffServ IPv4/IPv6 networks is significantly higher than scenarios 1 and 2. There may be couple of reasons; firstly, the offered traffic load is 50% higher than the assigned weight 5% (0.2 Mbps) for the AF13 traffic flows, which has highest drop probability when forwarding from the queue in which it's buffered. It is noted that PDV increases due to the congestion which leads to the packet to be waiting for long time in the queue. From the IPv6 protocol performance perspective, PDV in the IPv6 is relatively 7% higher than that of counterpart, IPv4. But packets are randomly dropped in the best-effort network as soon buffer is full.

In scenarios 5 and 6, Table 11 indicates that PDV for AF13 in IPv4 network shown in Figure 4(c) differs from approximately 0.0 $\mu$s to 18 $\mu$s, achieving an average PDV of about 10 $\mu$s. On the other hand, PDV for AF13 in IPv6 depicted in Figure 4(c) is varied from 0.0 $\mu$s to roughly 33 $\mu$s with an average 11 $\mu$s. By examining the obtain results of PDV with regard to the IPv6 protocol performance, IPv6 perceives 11% higher PDV than IPv4. In addition, from the Figure 4 and Table 11, it can be concluded that in the case of IP/DiffServ network, AF13 in the IPv6 network contributes to 7% higher PDV than that of IPv4 whereas PDV in the DiffServ/MPLS TE IPv6 network is 11% higher than that of IPv4.

Table 11. Summary statistics of PDV experienced by AF13 flows

| Scenarios | Min. [s] | Avg. [s] | Max. [s] | Std Dev [s] |
|---|---|---|---|---|
| Scenario 1 | 1,00E-10 | 3,52E-02 | 1,96E-01 | 4,61E-02 |
| Scenario 2 | 3,00E-10 | 6,46E-02 | 2,73E-01 | 7,64E-02 |
| Scenario 3 | 0,00E+00 | 4,76E+00 | 1,02E+01 | 4,04E+00 |
| Scenario 4 | 1,00E-10 | 5,12E+00 | 1,08E+01 | 4,42E+00 |
| Scenario 5 | 0,00E+00 | 1,02E-05 | 1,81E-05 | 6,60E-06 |
| Scenario 6 | 0,00E+00 | 1,16E-05 | 3,38E-05 | 7,68E-06 |

## 5.4 PDV Performance for AF41 Traffic Flows

Figure 5 demonstrates the comparable performance of the PDV under the different network scenarios measured for the video conferencing traffic flows generated at VC_Src4 and destined to VC_Dst4 through the IPv4 and IPv6 networks. The comparison against different network scenarios in terms of PDV is referred to Table 12, and observed in Figure 5.

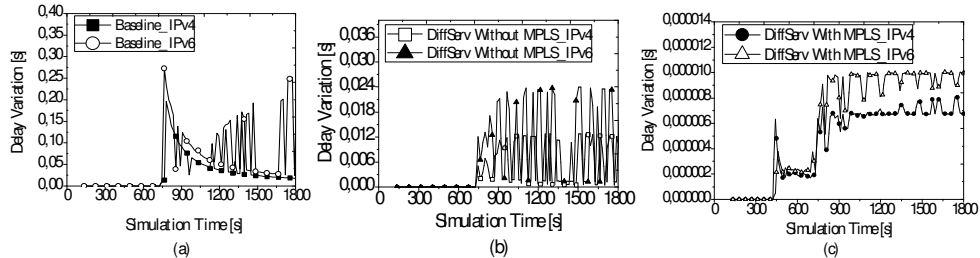

Figure 5. PDV experienced by AF41 flows for scenarios (a) 1-2 (b) 3-4 and (c) 5-6.

In scenarios 3 and 4, Table 12 points out that PDV for AF41 traffic flows in the IPv4 network (Figure 5(b)) differs from 0.1 ns to 13 ms achieving an average 4 ms. PDV for AF41 experienced in IPv6 network (Figure 5(b)) varies from 0.2 ns to 24 ms with an average 7 ms. From the IPv6 protocol performance perspective, PDV for AF41 in the IPv6 network is relatively 38% higher than counterpart, IPv4. Comparing to the baseline scenarios 1 and 2, one can easily see that PDV in IPv4/IPv6 networks has been significantly reduced. This is achieved for AF41 in IP/DiffServ network by setting up the highest priority with the weight 40% (1.6 Mbps) of the network capacity. Even though there is considerable difference in the PDV with respect to the Internet Protocol performance perspective.

In the scenarios 5 and 6, Table 12 indicates that the PDV for AF41 in the IPv4 network shown in Figure 5(c) differs from approximately 0.0 s to 8 µs, achieving an average PDV of about 4 µs. In contrast, PDV for AF41 in the IPv6 network depicted in Figure 5(c) (bottom middle side) is varied from 0.0 s to roughly 10 µs with an average 6 µs. By observing the simulation results of the PDV with regard to the IPv6 protocol performance, AF41 in IPv6 network perceives 25% higher PDV than IPv4 . In addition, from the Figure 5 and Table 12, it can be summarized that for AF41 traffic flows, PDV introduced in DiffServ IPv6 network is 38% higher than that of IPv4 while PDV in IPv6 network is also 25% higher than IPv4 when MPLS TE is introduced in the IP/DiffServ network.

Table 12. Summary statistics of PDV experienced by AF41 flows

| Scenarios  | Min. [s]  | Avg. [s] | Max. [s] | Std Dev [s] |
|------------|-----------|----------|----------|-------------|
| Scenario 1 | 1,00E-10  | 3,52E-02 | 1,96E-01 | 4,61E-02    |
| Scenario 2 | 3,00E-10  | 6,46E-02 | 2,73E-01 | 7,64E-02    |
| Scenario 3 | 1,00E-10  | 4,30E-03 | 1,30E-02 | 5,20E-03    |
| Scenario 4 | 2,00E-10  | 7,00E-03 | 2,40E-02 | 9,40E-03    |
| Scenario 5 | 0,00E+00  | 4,71E-06 | 8,09E-06 | 2,86E-06    |
| Scenario 6 | 0,00E+00  | 6,35E-06 | 1,00E-05 | 4,07E-06    |

**5.5 PDV Performance for AF42 Traffic Flows**

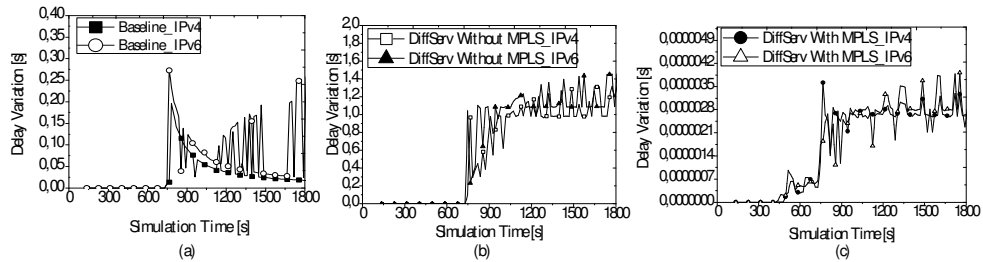

Figure 6. PDV experienced by AF42 flows for scenarios (a) 1-2 (b) 3-4 and (c) 5-6

Figure 6 shows the comparable performance of the PDV under the different network scenarios measured for the video conferencing traffic flows generated at VC_Src5 and destined to VC_Dst5 through the IPv4 and IPv6 networks.

The comparison against different network scenarios in terms of PDV is referred to Table 13, and observed in Figure 6. According to the simulation results shown in Table 13, by enabling the DiffServ in the baseline network scenarios, the obtained PDV for AF42 traffic flows in IPv4 network exhibited in Figure 6(b) is differed from 0.1 ns to 1.3 s reaching an average value 614 ms. At the same time, in IPv6 network, the maximum PDV is about 1.4 s, which is nearly 10% higher than that of IPv4. One can see that PDV for AF42 in the IPv4/IPv6 is higher than scenarios 1 and 2. This is because, the traffic prioritization set to medium and the respective assigned weight (15% of the link capacity) to the AF42, which can be defined as a trade-off. It is noteworthy to point out that a trade-off between mean PDV observed in Table 13 and less packet loss can be for the preferred and non-preferred flows. Concerning the IPv4/IPv6 protocol performance perspective, one can observe that PDV for AF42 in the IPv6 is relatively 11% higher than IPv4.

In the scenarios 5 and 6, Table 13 indicates the maximum PDV for AF42 in IPv4 network (Figure 6(c)) is approximately 3.7 $\mu$s, averaged 1.7 $\mu$s. On the other hand, the maximum perceived PDV for AF42 in IPv6 network (Figure 6(c)) is roughly 3.9 $\mu$s with an average 1.8 $\mu$s which is about 1.5% than that of IPv4. For the AF42, in comparison with scenarios 3 and 4, PDV is considerably lower in the IPv4/IPv6 networks. In addition, from the Figure 6 and Table 13, in the case of IP/DiffServ network, AF42 in IPv6 network contributes to 11% higher PDV than counterpart IPv4, and in conjunction with the DiffServ/MPLS TE, IPv6 is 1.5% higher than IPv4 as well.

Table 13. Summary statistics of PDV experienced by AF42 flows

| Scenarios  | Min. [s]  | Avg. [s] | Max. [s] | Std Dev [s] |
|------------|-----------|----------|----------|-------------|
| Scenario 1 | 1,00E-10  | 3,52E-02 | 1,96E-01 | 4,61E-02    |
| Scenario 2 | 3,00E-10  | 6,46E-02 | 2,73E-01 | 7,64E-02    |
| Scenario 3 | 1,00E-10  | 6,14E-01 | 1,32E+00 | 5,08E-01    |
| Scenario 4 | 1,00E-10  | 6,93E-01 | 1,46E+00 | 5,56E-01    |
| Scenario 5 | 0,00E+00  | 1,79E-06 | 3,62E-06 | 1,21E-06    |
| Scenario 6 | 0,00E+00  | 1,81E-06 | 3,91E-06 | 1,29E-06    |

**5.6 PDV Performance for AF43 Traffic Flows**

Figure 7 illustrates the comparable performance of the PDV under the different network scenarios measured for the video conferencing traffic flows which are generated at VC_Src6 and destined to VC_Dst6 through the IPv4 and IPv6 networks. The comparison against different network scenarios in terms of PDV is referred to Table 14, and observed in Figure 7.

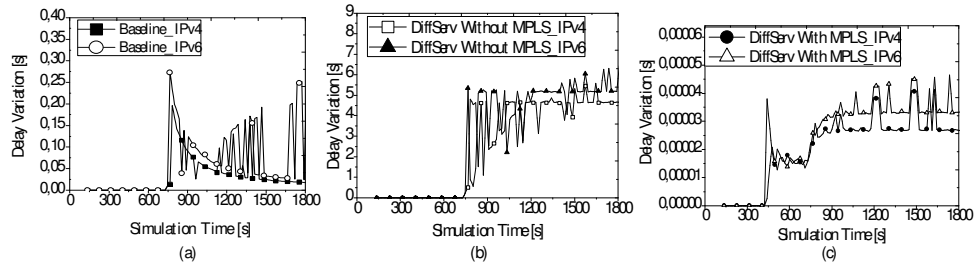

Figure 7. PDV experienced by AF43 flows for scenarios (a) 1-2 (b) 3-4 and (c) 5-6

In scenarios 3 and 4, PDV for AF43 in IPv4 network exhibited in Figure 7(b) is differed from 0.1 ns to 5.5 s achieving an average 2.63 s. At the same time, PDV experienced in DiffServ IPv6 network depicted in Figure 7(b) varies from 0.1 ns to 6.4 s with an average 3 s. From the IPv6 protocol performance perspective, PDV for AF43 in IPv6 is comparatively 11% higher than counterpart, IPv4. AF43 PDV in IPv4/IPv6 is quite higher than the PDV introduced in scenarios 1 and 2. This is due to the fact that AF43 is set to the lowest priority with 5% weight (0.2 Mbps) and the required bandwidth for this traffic is about 0.8 Mbps. In that context, for timely delivery of AF43 traffic depends on the ratio between the arrival and departure rates for the minimization of the queuing time introduced by the AF43 queue, as stated in [28]. In addition, the buffering time may also affect the on-time delivery of data packets, especially in real-time applications.

In scenarios 5 and 6, Table 14 indicates that the PDV for AF43 in IPv4 network shown in Figure 7(c) differs from approximately 0.0 s to 41 $\mu$s, achieving an average PDV of about 21 $\mu$s. In contrast, PDV in IPv6 network depicted in Figure 7(c) is varied from 0.0 s to roughly 46 $\mu$s with an average 24 $\mu$s. From the discussed analysis about PDV with regard to the IPv6 protocol performance, AF43 in DiffServ/MPLS IPv6 network perceives 11% higher PDV than that of IPv4. From the obtained results shown in Table 14, it can be concluded that on an average, PDV for AF43 in IPv6/DiffServ network is 11% higher than counterpart IPv4, and for the DiffServ/MPLS TE, PDV for AF43 in the IPv6 network is 11% higher than that IPv4.

Table 14. Summary statistics of PDV experienced by AF43 flows

| Scenarios  | Min. [s]  | Avg. [s]  | Max. [s]  | Std Dev [s] |
|------------|-----------|-----------|-----------|-------------|
| Scenario 1 | 1,00E-10  | 3,52E-02  | 1,96E-01  | 4,61E-02    |
| Scenario 2 | 3,00E-10  | 6,46E-02  | 2,73E-01  | 7,64E-02    |
| Scenario 3 | 0,00E+00  | 2,64E+00  | 5,58E+00  | 2,19E+00    |
| Scenario 4 | 1,00E-10  | 2,99E+00  | 6,43E+00  | 2,49E+00    |
| Scenario 5 | 0,00E+00  | 2,16E-05  | 4,16E-05  | 1,18E-05    |
| Scenario 6 | 0,00E+00  | 2,45E-05  | 4,64E-05  | 1,42E-05    |

### 5.7 Summary of PDV Performance

In Figure 8, the results of the comparison of PDV under 4 different scenarios are shown where Y axis represents PDV in seconds while X axis presents the various class-of-services (CoS). In the sense described above, the figure is to be interpreted in the following manner.

From the simulation results shown in Figure 8, it is clearly visible that the average PDV for AF11 flows in DiffServ/IPv6 network is found to be 12% higher than that of IPv4 while in DiffServ/MPLS, IPv6 remains 7% higher than IPv4. The average PDV for AF12 flows in DiffServ IPv6 network is appeared to be 10% higher PDV than that of IPv4, then again, for the DiffServ/MPLS TE, IPv6 is also 11% higher than IPv4. The average PDV experienced by AF13 traffic flows in DiffServ IPv6 network is 7% higher PDV than that of IPv4, whereas PDV in DiffServ/MPLS IPv6 network is considerably 11% higher than that of IPv4 as well. Now we turn to the average PDV for AF41 in the DiffServ IPv6 network is found to be 38% higher than that of IPv4, at the same time PDV in the DiffServ/MPLS IPv6 network is almost 25% higher than that of IPv4.

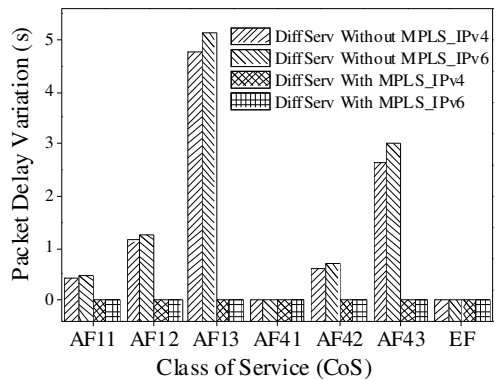

Figure 8. Average PDV against four different scenarios

Again the AF42 traffic flows in DiffServ IPv6 network, which suffers 11% higher PDV than that of IPv4 while AF42 in DiffServ/MPLS IPv6 network suffers 1.5% higher than that of IPv4. The average PDV for AF43 traffic flows in DiffServ IPv6 network is found to be roughly 11% higher than that of IPv4 whereas AF43 in DiffServ/MPLS IPv6 network is found to be 11% higher PDV than that of IPv4. Finally, EF in the DiffServ IPv6 network contributes to 8% higher PDV than that of IPv4, in contrast, EF flows in the DiffServ/MPLS IPv4 network contributes about 11% higher PDV than that of IPv6.

## 6. CONCLUSION

In this paper, we have evaluated the QoS performance of real-time applications in terms of PDV in IPv4/IPv6 networks. Six network scenarios have been simulated: Baseline IPv4 network, Baseline IPv6 network, DiffServ IPv4 Network, DiffServ IPv6 network, DiffServ/MPLS IPv4 network and DiffServ/MPLS IPv6 network. Comparative investigation of PDV performance was carried in four different network scenarios (e.g., DiffServ IPv4 Network, DiffServ IPv6 network, DiffServ/MPLS IPv4 network and DiffServ/MPLS IPv6 network). The research question was aimed to understand and investigate the performance of PDV for AF and EF PHBs vary from DiffServ/MPLS IPv4 network to DiffServ/MPLS IPv6 network. In our analysis, the average PDV for video traffic defined to the corresponding AF classes (i.e., AF11, AF12, AF13, AF41, AF42, and AF43) in the DiffServ IPv6 network was found to experience 5%~10% higher compared with the DiffServ IPv4 scenario. On the other hand, the average PDV in the DiffServ/MPLS IPv6 scenario was to be 7%~11% higher compared with the DiffServ/MPLS

IPv4 network. The average PDV for voice traffic corresponded to EF class in all the IPv6 network scenarios was found to virtually be indistinguishable compared with the IPv4 network scenarios.

In addition, our investigation shows that IPv6 experiences more PDV than their IPv4 counterpart.